\documentclass[aps,pra,twocolumn,showpacs,amsmath,amssym]{revtex4}

\usepackage{amssymb}
\usepackage{graphicx}
\usepackage{dcolumn}
\usepackage{verbatim}
\begin{document}

\title{Efficient numerical method to calculate three-tangle of mixed states}
\author{Kun Cao}
\affiliation{Key Laboratory of Quantum Information, University of
Science and Technology of China, Hefei, 230026, People's Republic of
China}

\author{Zheng-Wei Zhou}
\affiliation{Key Laboratory of Quantum Information, University of
Science and Technology of China, Hefei, 230026, People's Republic of
China}

\author{Guang-Can Guo}
\affiliation{Key Laboratory of Quantum Information, University of
Science and Technology of China, Hefei, 230026, People's Republic of
China}

\author{Lixin He
\footnote{Email address: helx@ustc.edu.cn} }
\affiliation{Key Laboratory of Quantum Information, University of
Science and Technology of China, Hefei, 230026, People's Republic of
China}

\date{\today }
\pacs{03.67.Mn, 03.65.Ud, 02.70.Ns, 02.70.Uu}

\begin{abstract}
We demonstrate an efficient numerical method to calculate
three-tangle of general mixed states. We construct a ``energy
function'' (target function) for the three-tangle of the mixed state
under certain constrains. The ``energy function'' (target function)
is then optimized via a replica exchange Monte Carlo method. We have
extensively tested the method for the examples with known analytical results, 
showing remarkable agreement. The
method can be applied to other optimization problems in quantum
information theory.

\end{abstract}

\maketitle 


One of challenges in present quantum information theory is the
quantification of quantum entanglement. The primary studies on
entanglement measures focus on bipartite systems. In bipartite
entanglement measures, formation of entanglement, presented by
Bennett et al., is very fundamental \cite{bennett96}. The
significance of this measure relies on not only be able to provide
an accurate boundary between separable states and entangled states
but also constructing a general formula for entanglement measures of
mixed states: once a measure for pure states is given, the
corresponding measure for mixed states can be obtained via the
convex-roof extension. This scenario can be even generalized to
quantify multipartite entanglement. The initial attempt is 3-tangle
of 3-qubit system, provided by Coffman et al \cite{coffman00}.
Entanglement measures beyond 3-qubit were also presented
\cite{Bai,Cai}. Although such kind of definition is straightforward,
the practical evaluation of the convex-roof extension is a difficult
mathematical problem. To date a general analytical method is known
only for the concurrence of two-qubit mixed states
\cite{uhlmann00,wooters98}.

Concurrence of mixed two qubits has been applied to study quantum
phase transitions. Most of studies show that, for general spin
$1/2$ lattice models, pairwise entanglement (or its functions)
depicted by concurrence of the nearest-neighbor two site has
special singularity at quantum critical points (see\cite{Vedral}
and its references). A natural generalization for this viewpoint
is that multipartite entanglement beyond two sites will reveal
more extensive characteristics on quantum phase transition.
However, as far as 3-tangle is concerned, since the convex roof is
obtained by minimizing the average tangle of a given mixed state
over all possible decompositions of pure states, there is no
general method to calculate the three-tangle of a mixed state. The
analytical method is only capable to study some particular three
qubit states\cite{lohmayer06,eltschka07,jung09,jung09a}.

In this work, we develop a method to calculate the three-tangle of
any mixed states by numerically minimize certain target functions.
The target functions are optimized by using replica exchange Monte
Carlo (MC) \cite{swendsen86,geyer_book} method. Replica exchange MC 
has been widely used in condensed matter physics to simulate the systems with
very rough ``energy surfaces'', such as spin glasses
\cite{marinari98}, protein systems and Lennard-Jones particles
\cite{sugita00,neirotti00} etc. 
The replica exchange MC method is much more effective to find the global minium of the
target functions than the conventional simulating annealing method. By
applying this method we have successfully obtained the three-tangle
of mixed (generalized) GHZ and W states. We also obtained the
three-tangle of GHZ state with white noise, whose density matrices
are of rank 8.


The concurrence $C(\psi_{AB})$ measures the biparticle entanglement
between particle A and B in a pure two-qubit state
$|\psi_{AB}\rangle$ . C is defined as
\begin{equation}
C = 2|\phi_{00}\phi_{11}-\phi_{01}\phi_{10}|.
\label{eq:C}
\end{equation}
in terms of the coefficients $\{\phi_{00}, \phi_{01}, \phi_{10},
\phi_{11}\}$of $|\psi_{AB}\rangle$ with respect to an orthonormal
basis.
The measure for three particle entanglement in a three-qubit state
$|\psi_{ABC}\rangle$ has been introduced in Ref.\cite{coffman00} as
three-tangle $\tau_3(\psi_{ABC})$. It can be expressed in terms of
the coefficients $\{\phi_{000},\phi_{001},...,\phi_{111}\}$.
\begin{equation}
\tau_3 = 4|d_1-2d_2+4d_3|
\end{equation}
\begin{equation}
d_1=\phi_{000}^2\phi_{111}^2+\phi_{001}^2\phi_{110}^2+\phi_{010}^2\phi_{101}^2+\phi_{100}^2\phi_{011}^2
\end{equation}
\begin{eqnarray}
d_2=\phi_{000}\phi_{111}\phi_{011}\phi_{100}+\phi_{000}\phi_{111}\phi_{101}\phi_{010}
\nonumber \\
+\phi_{000}\phi_{111}\phi_{110}\phi_{001}+\phi_{011}\phi_{100}\phi_{101}\phi_{010}
\nonumber\\
+\phi_{011}\phi_{100}\phi_{110}\phi_{001}+\phi_{101}\phi_{010}\phi_{110}\phi_{001}
\end{eqnarray}
\begin{equation}
d_3=\phi_{000}\phi_{110}\phi_{101}\phi_{011}+\phi_{111}\phi_{001}\phi_{010}\phi_{100}
\end{equation}
It can be shown that for any factorized state, the three-tangle vanishes.
For the GHZ state,
\begin{equation}
|GHZ\rangle=\frac{1}{\sqrt{2}}(|000\rangle+|111\rangle)
\end{equation}
$\tau_3(GHZ)$=1. It also appears a class of
entangled three-qubit states represented by $|W\rangle$ state with
$\tau_3$ vanished, i.e.,
\begin{equation}
|W\rangle  = \frac{1}{\sqrt{3}}(|100\rangle+|010\rangle+|001\rangle)
\end{equation}
The three-tangle of mixed state can be obtained via convex-roof
extension\cite{uhlmann00,bennett96,benatti96}.
Suppose the density matrix of a mixed state $\rho$
can be decomposed into sum of some pure states $\pi_i$, i.e.,
\begin{equation}
\rho = \sum_{i}p_i\pi_i \, ,
\end{equation}
where,
\begin{equation}
\pi_i ={|\Psi_i\rangle\langle\Psi_i|}\, .
\label{eq:DenMat}
\end{equation}
$|\Psi_i\rangle$ is the wave function of a normalized pure state.
The three-tangle of the mixed state is defined as the average pure-state concurrence
minimized over all possible decompositions.
\begin{equation}
\tau_3(\rho) = \min \sum_i p_i \tau_3(\pi_i) \, .
\label{eq:mixtau3}
\end{equation}
%


It is very difficult to develop an universal analytical method to
calculate the three-tangle of mixed three-qubit state. So far, there
are very limited examples of mixed states whose three-tangle have
been obtained
analytically\cite{lohmayer06,eltschka07,jung09,jung09a}.
Alternatively, one could resort to the numerical methods. To get the
three-tangle of mixed states is a constrained minimization problem.
The pure state wavefunction can be expanded on the orthonormal
three-particle basis, i.e., $|\Psi_i\rangle =
\sum_{\alpha}c_{i\alpha}|\Phi_{\alpha}\rangle$.
The first constrain is,
\begin{equation}
\sum_i p_i\, c_{i,\alpha} c^*_{i,\beta} = \rho_{\alpha\beta}.
\label{eq:con_R2}
\end{equation}
where $\rho_{\alpha\beta}$ is the density matrix in the basis of  $
\{| \Phi_{\alpha} \rangle \}$. Furthermore,
the coefficients have to satisfy the
normalization conditions,
\begin{equation}
\sum_\alpha |c_{i\alpha}|^2=1, \, {\rm and} \,
\sum_{i} p_i=1, p_i \geq 0\, .
\label{eq:constrain2}
\end{equation}
The above constrains can be enforced in the simulation by
explicitly applying normalization factors at each MC step,
whereas the constrain Eq. \ref{eq:con_R2} can be enforced
via a penalty function.
The ``energy function'' (target function) reads,
\begin{equation}
E(\{ p_i,\Psi_i \})= \sum_i p_i\tau_3(\Psi_i) +\kappa R^2 \, ,
\label{eq:target}
\end{equation}
where $\{ p_i,\Psi_i \}$ realizing the mixed state as given in
Eq.\ref{eq:DenMat}.
 $R^2$ is the residual between the searched density
matrix and the target density matrix $\rho$, defined as
\begin{equation}
R^2= \sum_{\alpha=1}^{N_c} \sum_{\beta=1}^{N_c}
\left(\sum_{i=1}^{N_p} p_i\, c_{i,\alpha} c^*_{i,\beta} -
\rho_{\alpha\beta} \right)^2 \, .
\end{equation}
$\kappa$ is a large constant, to ensure $R^2 \sim$0. $N_c=8$ is the dimension of
three-particle basis set. $N_p$ is the number of partition in the
decomposition. We can easily see that $N_p$ determines
the number of variables $N_v=N_p+2N_cN_p$.


In Eq.\ref{eq:target}, $\kappa$ should be large enough to ensure a
small residual $R^2$ to numerically satisfy the constrain given in
Eq.\ref{eq:con_R2}. However, large $\kappa$ will make the ``energy''
surface of $E$ very rough, i.e., $E(\{ p_i,\Psi_i \})$ has many
local minima separated by large ``energy'' barriers, which causes
great difficulties in finding the global minima of three tangle
using traditional constrained simulated annealing (CSA) method,
because it tends to be trapped at some local minima. Here, we adopt
the replica exchange (also known as parallel tempering \cite{young_book}) Monte
Carlo method \cite{swendsen86} which simulates $M$ replicas
simultaneously each at a different temperature $\beta_0=1/T_{\rm
max} <\beta_1<...<\beta_{n-2}<\beta_{M-1}=1/T_{\rm min}$ covering a
range of interest. Each replica runs independently, except that
after certain steps, the configurations can be exchanged between
neighboring temperatures, according to the Metropolis criterion,
\begin{equation}
w=\left \{ \begin{array}{ll}
1 \quad & \Delta H < 0 \\
e^{-\Delta H}  & \mbox {\rm otherwise}\;,  \end{array} \right.
\label{eq:repex3}
\end{equation}
where $\Delta H= -(\beta_{i}-\beta_{i-1}) (E_i-E_{i-1})$, in which
$E_i$  and $E_{i-1}$ are the energy of the $i$-th and $i$-1-th
replica. During the exchange, the detailed balance in canonical
ensemble is satisfied.  Importantly, the inclusion of high-$T$
configurations ensures that the lower-$T$ systems can access a broad
phase space and avoid becoming trapped at local minima. The replica
temperatures are adjust so that the exchange rate between the
replicas are all about 20\%\cite{kone05}. while keep the highest temperature
$\beta_0=1/T_{\rm max}$ and lowest temperature $\beta_{M-1}=1/T_{\rm
min}$ fixed. This can be done by adaptively adjusting the
temperature of each replica using a recursion method proposed by
Berg\cite{berg_book}. We slightly revised the the algorithms
to give better performance. Suppose, in the $n$-th iteration, the
temperature of the $i$-th replica is $\beta_{i}^{n}$ and the
acceptance rate between $\beta_{i}^{n}$ and $\beta_{i-1}^{n}$ is
$a_{i}^{r,n}$. In the $n$+1-th iteration, $\beta_{i}^{n+1}$ can be
updated as,
\begin{equation}
\beta_{i}^{n+1}=\beta_{i-1}^{n+1}+(1-c+ca_i^{n})(\beta_i^{n}-\beta_{i-1}^n)
\, , \label{eq:recursion}
\end{equation}
where,
\begin{equation}
a_i^{n}=\frac{a_{i}^{r,n}(\beta_{M-1}^{n}-\beta_{0}^{n})}
{\sum_{i=1}^{M-1}{a_{i}^{r,n}(\beta_i^{n}-\beta_{i-1}^n)}} \, ,
\end{equation}
$c$ is an empirical parameter that can be adjusted to accelerate the
convergency. In the present case, we found that by choosing
proper $c<1$ (in this case c=0.7) the recursion scheme get a suitable
$\beta$ distribution within tens of iterations. We then fixed the
replica temperatures for the simulations.


Based on convex-roof extension, Lohmayer \textit{et al.} have
provided a complete analysis of mixed three-qubit states composed of
a GHZ state and a W state, obtaining the optimal decomposition and
convex-roof for three-tangle\cite{lohmayer06}. Eltschka \textit{et
al} generalized this method to treat the three-tangle of mixed
three-qubit states composed of a generalized GHZ state and a
generalized W state. The explicit expressions for mixed-state three-
tangle and the corresponding optimal decomposition for the more
general form were also presented\cite{eltschka07}.
We first test our scheme by comparing with the available analytical results.
We construct a series of target functions based on the density matrix given in
Refs.~\onlinecite{lohmayer06,eltschka07} as were described in the previous
sections. Replica exchange MC is then performed to find the
three-tangle and optimal decomposition which are compared with the
analytical results.

In all our minimization process, the temperature range is chosen
between $T_{\rm max}$=100 and $T_{\rm min}$=10$^{-6}$, such that
$T_{\rm max}$ is high enough to help the low temperature replica to
avoid being trapped in the local minimum and $T_{\rm min}$ is low
enough to give the final minimized value an accuracy of $10^{-6}$ theoretically.
With fixed temperature range, the suitable number of replicas is
determined by the corresponding density matrix and weight factor
$\kappa$. Usually 150 replicas are enough. In our simulations, $\kappa$
is adjusted to be $10^4-10^7$ to keep the final $R^2 < 10^{-10}$
which is considered accurate enough in identifying the three-tangle
of mixed state. In each simulation, the number of partitions $N_p$
is fixed. We then increase $N_p$ until the calculated three-tangle
converges.

\begin{figure}
\centering
\includegraphics[width=2.8in]{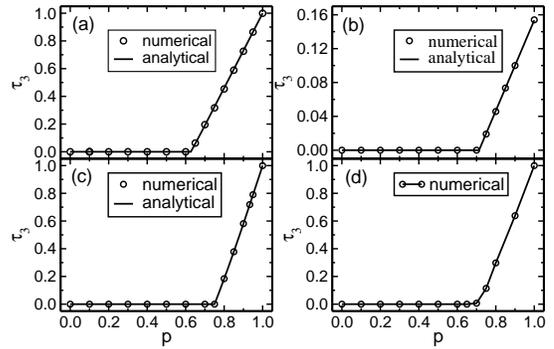}
\caption {The three-tangle calculated with replica exchange compared to analytical results,
for (a) the mixed GHZ and W states, 
(b) the mixed generalized GHZ and generalized W states,
(c) the mixed GHZ state, W state and flipped W states with n=2,
(d) the mixed GHZ state and white noise states.}
\label{fig:tau3}
\end{figure}

Let us first look at the mixed GHZ and W states,
The density matrices of the states are,
\begin{equation}
\rho(p)=p\,\pi_{GHZ}+(1-p)\,\pi_{W} \, ,
\end{equation}
where $\pi_{GHZ}=|GHZ\rangle\langle GHZ|$ and
$\pi_{W}=|W\rangle\langle W|$ are the density matrices of
GHZ and W states respectively\cite{lohmayer06}.
It is known that $\tau_3(GHZ)$=1 and $\tau_3(W)$=0. According
to Caratheodory's theorem, $N_p$=4 is enough to get the optimized
$\tau_3$ of a mixed states of rank 2\cite{lohmayer06}. We calculate $\tau_3$
by using $N_p$=4 - 8.
Indeed,
we find in our simulations, that $N_p$=4 is enough to give the optimal values. The
results are shown Fig.\ref{fig:tau3}(a), which
depicts the numerically calculated $\tau_3$ of the
mixed GHZ and W states (open circles), compared to the analytical
values (solid line)\cite{lohmayer06}. As one can see that the
numerical and analytical results are in excellent agreement.

We then calculate $\tau_3$ for the mixtures of generalized
GHZ and generalized W states\cite{eltschka07}, whose density matrices
are,
\begin{equation}
\rho(p)=p\,|gGHZ\rangle\langle gGHZ|+(1-p)\,|gW\rangle \langle gW| \, ,
\end{equation}
where the density matrix of a generalized GHZ state is
\begin{equation}
|gGHZ\rangle = a|000\rangle+ b|111\rangle \, ,
\end{equation}
and density matrix of a generalized W state is
\begin{equation}
|gW\rangle = c|001\rangle+ d|010\rangle+f|100\rangle \, .
\end{equation}
The coefficients must satisfy the normalization condition
$|a|^2+|b|^2$=1, and $|c|^2+|d|^2+|f|^2$=1. For the pure generalized
GHZ and W states, it is easy to calculate that $\tau_3(gW)$=0, and
$\tau_3(gGHZ)=4|a^2b^2|$.

The calculated $\tau_3$ of the mixed states are shown in Fig.\ref{fig:tau3}(b)
for $a$=0.2, $c$=0.2, $d$=0.2. As we see the results
are in excellent agreement with
the analytical results. Again we find that $N_p$=4 is enough to
the optimal results.

In the above tests, we demonstrate that our method is very effective
to deal with the three-tangle of rank-2 mixed states.
Recently, Jung at al. has obtained the analytical results of
the three-tangle of the mixed states of
GHZ, W, and flipped-W states\cite{jung09}. These states are of rank 3.
The flipped-W states is defined as
\begin{equation}
|\tilde{w}\rangle = \frac{1}{\sqrt{3}}(|110\rangle +|101\rangle +
|011 \rangle) \, .
\end{equation}
The density matrix of mixture of GHZ, W, and flipped-W states are given by,
\begin{equation}
\rho(p,q)=p\,|GHZ\rangle\langle GHZ|+q\,|W\rangle \langle W|
+(1-p-q)\,|\tilde{W}\rangle \langle \tilde{W}|\, ,
\end{equation}
where $q$ was defined as $q=1-p/n$, n is a positive number.
We then compared our numerical results to the analytical values.
The results for the case of n=2 are shown
in Fig.\ref{fig:tau3}(c).
The numerical results are also in good agreement with the analytical results. 
We further calculate three-tangle of the mixed states which are of rank 4,
given in Ref.\onlinecite{jung09a}, obtaining the same accuracy compared to the
analytical results.

\begin{table}
\caption{Numerically calculated $\tau_3$ of mixed GHZ, W and flipped-W
states for n=2 compared to the analytical results given in 
Ref.~\onlinecite{jung09}.}
\begin{center}
\begin{tabular}{cccccccc}
\hline\hline
    p      & 0.2 & 0.75 & 0.8 & 0.85 & 0.9 & 0.933 & 0.95  \\
\hline
analytical &  0     & 0      & 0.1835 &0.3775 & 0.5805 & 0.7182 & 0.7897 \\
numerical  & 5.7e-6 & 2e-5 & 0.1836 &0.3781 & 0.5811 & 0.7186 & 0.7901 \\
\hline \hline
\end{tabular}
\label{tab:gww1}
\end{center}
\end{table}

To give an illustration of the accuracy, the numerical values of mixture of
GHZ, W and flipped-W
state with n=2 are also listed in Table \ref{tab:gww1}.
We can see that all the results are within an accuracy of $\sim 10^{-4}$.
The accuracy can easily be improved by
increasing $\kappa$ and decreasing $T_{\rm  min}$.
For the example given above, in the case of p=0.8,
if we reduce $T_{\rm  min}$ to $10^{-10}$ and use $\kappa=10^{10}$, we obtain
$\tau_3$=0.18349906 compared to the analytical value 0.18349861
in an accuracy $\sim 10^{-7}$. In principle,
our numerical method can achieve arbitrary accuracy.
We have also tried to optimize Eq. \ref{eq:target} via traditional CSA method.
We find the performance of the traditional method is very bad, because
it can easily be trapped at some local minima.

\begin{figure}
\centering
\includegraphics[width=2.4in]{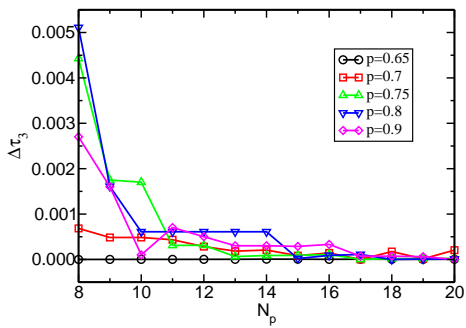}
\caption {(Color online) The $N_p$ dependent $\Delta\tau_3$ calculated with
  replica exchange for
density matrix constructed by GHZ state with white noise,
where $\Delta\tau_3=\tau_3-\tau_3^{min}$ with $\tau_3^{min}$ the minimum three-tangle for each $p$ value.}
\label{fig:converge}
\end{figure}

Having demonstrated the ability of the method,
we then calculate the $\tau_3$ of the GHZ states
mixed with white noise, which does not have analytical solutions so far.
The density matrices of the states are
\begin{equation}
\rho(p)=p\,\pi_{GHZ}+\frac{1-p}{8}\,\pi_E\, ,
\end{equation}
where $\pi_E$ is a unit matrix. In this case, the density matrices
are of rank 8. We need at least $N_p$=8 to ensure $R^2 \sim$ 0.  In
practice, we have calculated $\tau_3$ up to $N_p$=20. We find that
use $N_p$=15 can converge $\tau_3$ to less than 10$^{-3}$ for this
problem. (See Fig. \ref{fig:converge}). The calculated results are
shown in Fig.\ref{fig:tau3}(d). 
From Fig. \ref{fig:tau3}(d), we see that
$\tau_3$ of the white noise mixed GHZ states is zero for $p$ less
than about 0.7, then increase almost linearly to 1.0 as $p$
increases. The behavior of $\tau_3$ looks similar to that of mixed
GHZ and W states.
 
It is straight
forward to apply the method to any mixed states.
It therefore opens up a way to study the
multiparticle entanglement effects in many important systems, e.g.,
in systems with quantum phase transitions, which was impossible before,  
because the analytical solution to the tree-tangle of a general mixed
state is not available.

To conclude, we have demonstrate an efficient numerical method to calculate the
three-tangle of general mixed states. We construct a ``energy function'' (target function)
for the three-tangle of the mixed state under certain constrains.
The ``energy function'' (target function) is then optimized via a replica
exchange Monte Carlo method.
We have tested the method for the examples with known
analytical results, showing remarkable agreement.
We further
calculate the three-tangle of GHZ states mixed with white noise.
The work opens a way to study three-tangle of general mixed states, and can be
generalized to solve many other
optimization problems in quantum information theory.

L.H. acknowledges the support from the Chinese
National Fundamental Research Program 2006CB921900, the Innovation
funds and ``Hundreds of Talents'' program from Chinese Academy of
Sciences, and National Natural Science Foundation of China, Grant
No. 10674124. W.Z were supported by National Natural Science
Foundation of China, Grant No. 10874170 and K.C.Wong Education
Foundation, Hong Kong.


\end{document}